**Accelerating Hydrodynamic Fabrication of Microstructures using Deep Neural Networks**

Nicholus R. Clinkinbeard [a], Reza Montazami [a], and Nicole N. Hashemi [a,*]

[a] Department of Mechanical Engineering, Iowa State University, Ames, IA, 50011, USA

* Corresponding author: nastaran@iastate.edu,

**Abstract**

Manufacturing of microstructures using a microfluidic device is a largely empirical effort due to the multi-physical nature of the fabrication process. As such, models are desired that will predict microstructure performance characteristics (e.g., size, porosity, and stiffness) based on known inputs, such as sheath and core fluid flow rates. Potentially more useful is the prospect of inputting desired performance characteristics into a design model to extract appropriate manufacturing parameters. In this study, we demonstrate that deep neural networks (DNNs) trained with sparse datasets augmented by synthetic data can produce accurate predictive and design models. For our predictive model with known sheath and core flow rates and bath solution percentage, calculated solid microfiber dimensions are shown to be greater than 95% accurate, with porosity and Young's modulus exhibiting greater than 90% accuracy for a majority of conditions. Likewise, the design model is able to recover sheath and core flow rates with 95% accuracy when provided values for microfiber dimensions, porosity, and Young's modulus. As a result, DNN-based modeling of the microfiber fabrication process demonstrates high potential for reducing time to manufacture of microstructures with desired characteristics.

# 1 Introduction

In recent years, microfluidics has increasingly proven itself an invaluable approach to the fabrication of microstructures supporting bottom-up tissue engineering strategies, facilitating the development and assembly of living building blocks [1], [2]. However, in order for microfluidics to be a viable solution to generation of microfibers on a large scale that serve research and industrial applications, a methodology is necessary that will allow selection of manufacturing parameters not based solely on trial-and-error. Specifically, the state of the art must move toward predictive modeling using existing empirical data and physical theory.

Due to their proclivity for determining complex relationships among a number of parameters in a dataset [3], artificial intelligence (AI) and machine learning (ML) techniques in particular have become commonplace for predicting behavior of physical phenomena [4] - [7]. Unfortunately, in the presence of sparse data, generalization for models developed with such techniques may not be feasible [8] - [10]. To circumvent the issue of having too little empirical information, synthesis of reliable data is an active area of study for several fields, including medical and even nonphysical topics, such as data privacy [11], [12]. Oftentimes, particularly in image processing, minor alterations are made to original sources in order to increase the amount of available data [13].

In this study, a deep neural network approach is applied to the fabrication process for solid microfibers manufactured under the following conditions described by McNamara et. al [14]: i) microfluidic chip with two sheath and one core solution inlet, ii) core solution of 6% alginate dissolved in water, iii) sheath solution comprising 0.5% calcium chloride dihydrate ($CaCl_2$-$H_2O$) and 5% poly ethylene glycol (PEG), iv) bath solutions of 0% and 5% CaCl2-2H20, and v) polymerization via chemical crosslinking.

Two objectives are in view: (1) prediction of fiber features based on given manufacturing parameters, and the converse, (2) determination of manufacturing parameters that will produce desired fiber features. The first of these uses a DNN to develop a model for enhancing accuracy of system performance predictions, while the second uses a similar approach to generate a model that determines a narrow subset of tests to conduct by providing recommended manufacturing parameter values. Due to the scarcity of available data for training and testing a DNN model, datasets are synthesized using the statistical properties of baseline data collected experimentally and presented by McNamara et. al [14].

The significance of this study is also twofold. However, while the first objective is valuable in an effort to understand how different parameters affect resulting microfiber features, the primary benefit lies with the second objective. Successful implementation of a neural network to design manufacturing parameters reduces the amount of trial and error required to produce a viable microfiber, which saves both time and resources.

## 2 Methods

To set the stage for development of suitable microfiber generation models, it is important to first define the desired outputs of such models and formulate an approach (or approaches) to achieve these outputs. To accomplish this, we describe the following activities:

- identification of a suitable goal-oriented modeling approach and
- implementation of a deep neural network to generate predictive and design models.

### 2.1 Modeling Approach

A methodology that appropriately addresses input parameters, output parameters, and intermediary concerns is critical to the development of models that can either predict microfiber characteristics given a set of manufacturing parameters or to design manufacturing parameters based on desired microfiber characteristics. Due to the importance of both—as well as their distinction—time is expended to provide definitions for the frameworks developed and implemented herein, beginning with performance parameters and the design space.

### 2.1.1 Performance Parameters and Design Space

Crucial to the proper formulation of an AI-based microfiber fabrication model is an appropriate understanding of various related parameters and their functions. As such, we have divided these into four main categories: *static design*, *dynamic design*, *performance*, and *calculated*. While these concepts in and of themselves represent nothing new to the field of data science, their categorization as such within this section is important for understanding the subsequent modeling strategies.

To begin, *static design parameters* are those inputs that are not changeable for the system at hand but are an indispensable part of the process. For the case of microfiber fabrication, this primarily comprises microfluidic chip geometry. While, technically, alternate microfluidic chip designs are possible, we have chosen to study the existing system described in several previous studies [14] - [17] and shown in Figure 1. Therefore, while features of the chip—such as cross-sectional flow area and chevron arrangement—are important considerations, they are not variable throughout this evaluation. As a result, since we have limited the investigation to a data-driven approach without additional physics-based formulations, static design parameters associated with the microfluidic chip are not explicitly used in the DNN-based modeling process for this phase of the study. Additional static design parameters of note for this effort are the core and sheath solutions. The core solution

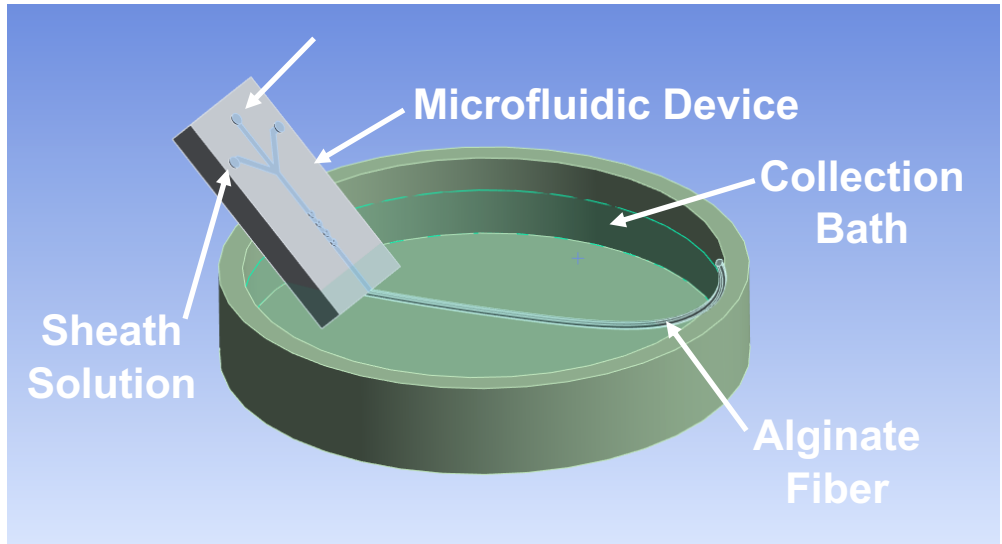

**Figure 1.** Schematic of the microfluidic fiber fabrication process for solid alginate microfibers. Note that sheath and core solutions are pumped into the microfluidic device using precisely-controlled and steady flowrates.

In contrast to static parameters, *dynamic design parameters* are those inputs to a process that can be changed to effect a desired outcome. For microfiber fabrication, these can include fluid selection, sheath and core flow rates, solution concentrations, and polymerization methodology, among others. Further derived from these basic parameters are relevant fluid properties, such as density and viscosity for the different solutions. Again, since we are limiting the DNN-based model development for this study to data processing at the exclusion of physics-based information, we limit dynamic design parameters to sheath and core flow rates, as well as bath solution concentration.

Third, *performance parameters* are those that describe how the system works and are thus directly related to experimental outcomes. These are features of the final fabricated microfibers that are of particular importance to their ultimate research and industrial use. Of specific interest to this study are fiber cross-sectional dimensions, Young's modulus, and porosity. Although typically evaluated as an output of a model comprising static and dynamic inputs, a process that uses performance parameters as an input for determining manufacturing inputs (i.e., dynamic design parameters) would greatly accelerate the microfiber fabrication design and development process.

One thing of importance to note is that these definitions do not relegate any characteristic as specifically an input or output, at least with respect to a modeling approach.[1] Whether a parameter functions as an input or an output is directly related to the type of model to which it belongs—either one that is predictive, or one that is intended as a design tool. These are defined and discussed in the following section as drivers for the introduction of two basic modeling architectures.

### 2.1.2 Basic Model Architectures

To advance the development of a model-based microfiber fabrication process, we now discuss modeling architecture. At this point we are agnostic to the means of modeling, i.e, the finite element method (FEM) and computational fluid dynamics (CFD), artificial intelligence, reduced-order models (ROMs), etc. However, the information provided here directly informs our emergent approach.

#### 2.1.2.1 Predictive Model

The purpose of a predictive model is to forecast system performance given a set of known input parameters. Such input parameters are subjected to a mathematical process, which in turn predicts outputs. Although oversimplified here for brevity, this methodology is pervasive in numerical modeling practices, such as the finite element method or computational fluid dynamics. The investigator supplies known static and dynamic design inputs—geometry, material and fluid properties, loads, initial and boundary conditions, etc.—into the model, which thereby predicts specific performance parameters, such as displacement, velocity, force, stress, etc., using known physics-based principals or empirically-derived formulae synthesized with the appropriate mathematics. In the case of FEA, CFD, and other numerical techniques, the process includes discretization and algebraic approximation of the governing equations.

---

[1] As opposed to an experimental approach, where the manufacturing parameters will always be the inputs and the performance parameters the outputs.

Within the context of microfiber fabrication using microfluidic devices, a predictive model is used to narrow experimental options to those that are believed to produce a desired set of microfiber properties. Specifically, the goal of a microfiber predictive model is to input sheath and core fluid properties and flow rates with consideration for microfluidic chip properties and bath solution information in order to determine microfiber performance characteristics, such as geometry, porosity, Young's modulus, and strength. While conceptually, the use of predictive models is similar to experimentation (i.e., modeling and testing both subject inputs to a process and generate observable results), it is potentially a less expensive endeavor, both with respect to time (running simulations is typically faster than conducting experiments) and resources (simulations can reduce the amount of raw material, equipment, and labor costs for a study). A basic flow diagram of this predictive modeling process is shown in Figure 2.

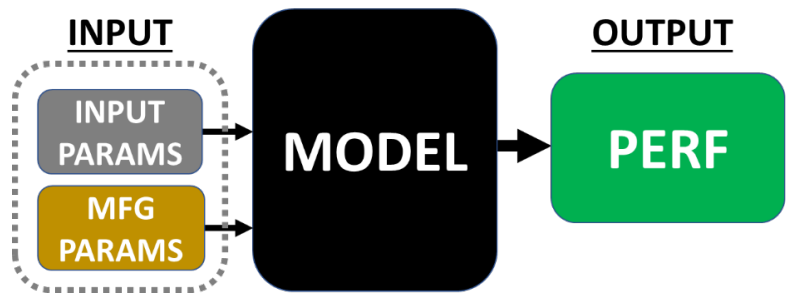

**Figure 2.** This basic flow diagram of a *predictive model* is used to take assumed input parameters (fluid properties, flow rates, etc.) and extract performance of the resulting microfiber (geometry, strength, porosity, modulus). The process is non-optimal for design in the sense that it mimics experimentation, relying heavily on trial and error and engineering judgment.

The downside of a predictive model approach for the development of microfibers with desired characteristics is that it involves a potentially significant amount of trial and error, similar to experimentation. The investigator must preselect fluids for the process, as well as flow rates and other experimental parameters. The modeling process then produces resulting fiber characteristics that may or may not match desired properties. The solution to this issue is to reorganize the model such that desired fiber characteristics serve as inputs, which we next discuss.

### 2.1.2.2 Design Model

The purpose of a design model is to *begin* with desired system performance and extract appropriate input parameters to achieve these characteristics. Such an approach is essentially an optimization problem whereby a certain performance criterion (or criteria) is selected and the design space is explored to determine appropriate and practical input parameters that will lead to preferred performance. Stated another way, the inputs to the model are the desired fiber characteristics, while the outputs are the manufacturing parameters. The advantage to this methodology is if the model is accurate, we are able to directly apply the manufacturing parameters experimentally and should expect fiber characteristics that largely accord with our desires.

Figure 3 shows a simple block diagram of the microfiber fabrication modeling process using a design approach.

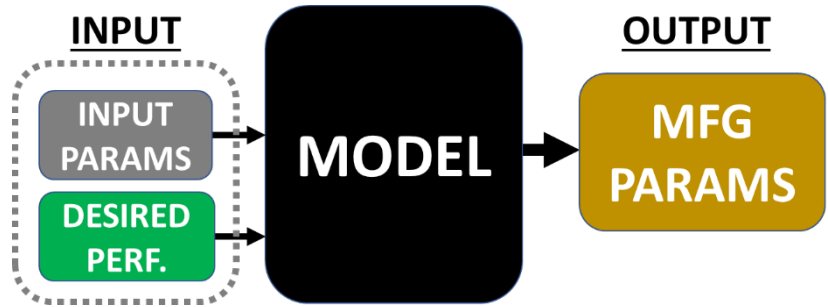

**Figure 3.** This basic flow diagram of a *design model* is used to take desired microfiber features (geometry, strength, porosity, modulus) (fluid properties, flow rates, etc.) and extract manufacturing parameters that lead to such results (fluid properties, flow rates, etc.). The process is directed toward first-time experimentation success by providing optimal manufacturing parameters to be used with the physical fabrication setup.

### 2.2 Model Development: Deep Neural Network

Thus far we have demonstrated the need for and discussed some of the advantages of developing an accurate predictive or design model to reduce the amount of experimentation required to produce viable microfibers using a microfluidic chip. What we have not done is to specify the type of model, i.e., (FEA, ROM, AI, etc.). While a predictive model allows us to conduct "what if" studies by parameter variation, the advantage to having an accurate design model should be obvious—we are effectively able to remove the majority of trial and error from both the modeling process and subsequent experimentation. However, the key to success with either approach is the creation of an accurate model, which is a nontrivial task. At this point we discuss the models developed for our study, which are based on a deep neural network architecture.

### 2.2.1 Artificial Deep Network Architecture

Use of artificial intelligence within the microfiber generation process provides an opportunity to use historical test data—albeit sparse—to aid in the prediction of fiber performance and, more importantly, design *for* performance. To demonstrate the former, we modify the basic flow diagram of Figure 2 to specify artificial intelligence as the modeling approach, as shown in layers with several neurons each; and one output layer with four features (fiber length, $w_f$, fiber width, $l_f$, fiber porosity, $\Phi_f$, and fiber Young's Modulus of Elasticity, $E_F$.)

Figure 4(a). The method selected for generating a predictive model based on AI is a deep artificial neural network. A schematic of the DNN used in this process is shown in Figure 4(b). Note that the predictive model specifically uses as input parameters mass flow rate of the sheath and core fluids ($\dot{V}_S$ and $\dot{V}_C$, respectively) along with percentage bath solution ($C$). Resulting performance parameters are fiber cross-sectional dimensions (length, $l_f$ and width, $w_f$), porosity ($\Phi_f$), and Young's Modulus ($E_f$).

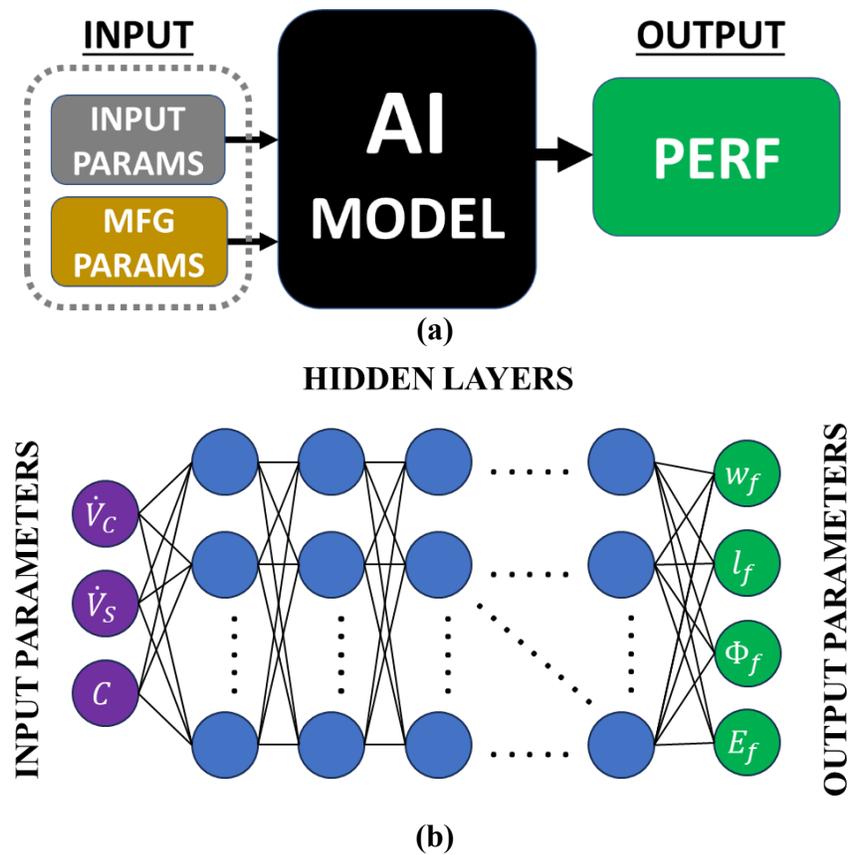

**Figure 4.** (a) Basic flow diagram for the microfiber *predictive model* process. Although a viable part of the overall modeling methodology, the inclusion of reduced-order-of-magnitude models is deferred to the next phase of this study. (b) Basic deep neural network used to predict performance parameters for the microfiber fabrication process. Note the following characteristics of this DNN: three input features (core flow rate, $\dot{V}_C$, sheath flow rate, $\dot{V}_C$, and bath concentration, $C$); hidden

To demonstrate the fiber design model, a modification of Figure 3 showing implementation of artificial intelligence to generate a suitable model is provided in Figure 5(a). Once again, the approach selected as the AI-based model is a deep artificial neural network, for which a schematic is shown in Figure 5(b). Note that the input and output parameters are exactly swapped from the predictive model.

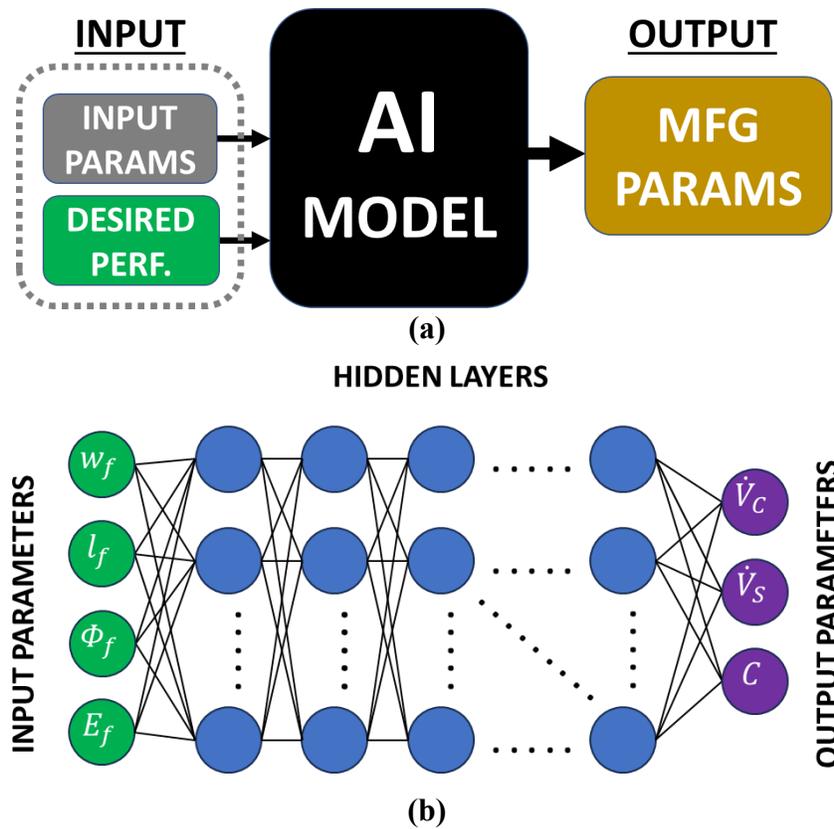

**Figure 5.** (a) Basic flow diagram for the microfiber fabrication *design model* process. Although a viable part of the overall modeling methodology, the inclusion of reduced-order-of-magnitude models is deferred to the next phase of this study. (b) Basic deep neural network used to design manufacturing parameters for the microfiber fabrication process. Note the following characteristics of this DNN: four input features (fiber length, $w_f$, fiber width, $l_f$, fiber porosity, $\Phi_f$, and fiber Young's Modulus of Elasticity, $E_F$); hidden layers with several neurons each; and one output layer with four parameters (core flow rate, $\dot{V}_C$, sheath flow rate, $\dot{V}_C$, and bath concentration, $C$).

A major disadvantage of the approach taken in this study is that it tends to require large datasets to accurately train a DNN. To alleviate this issue, synthetic data were generated based on the original experimental results, which is addressed in the following section.

### 2.2.2 Experimental Basis

Experimental data used as the basis for implementation of AI-based predictive and design models are for fabrication of alginate microfibers with solid cross section, as derived from McNamarra et. al [14].[2] That investigation evaluated solid fiber characteristics for the flow rate ratios and bath $CACL_2$ concentration values shown in Table 1. For the study at hand, this results in the following input (design) parameters: sheath flow rate, core flow rate, and bath solution.

**Table 1.** Solid alginate microfiber manufacturing parameters (taken from McNamarra et. al [14].

| Flow Rate Ratio (Sheath:Core) | Bath Solution Concentration |
|---|---|
| 100:10 | 0% |
| 125:10 | 5% |
| 125:15 | |

Performance parameters extracted from [14] are manufactured fiber porosity, Young's modulus, and fiber dimensions. Specifically, length and width were estimated assuming a rectangular fiber cross-section. Although an estimate, an example of length and width definition is shown in Figure 6.

---

[2] It must be noted that aspects of the data ignored from the prior study are cell embedment and viability.

Table 2 summarizes the mean and standard deviation of measured features for each microfiber input and performance parameter.

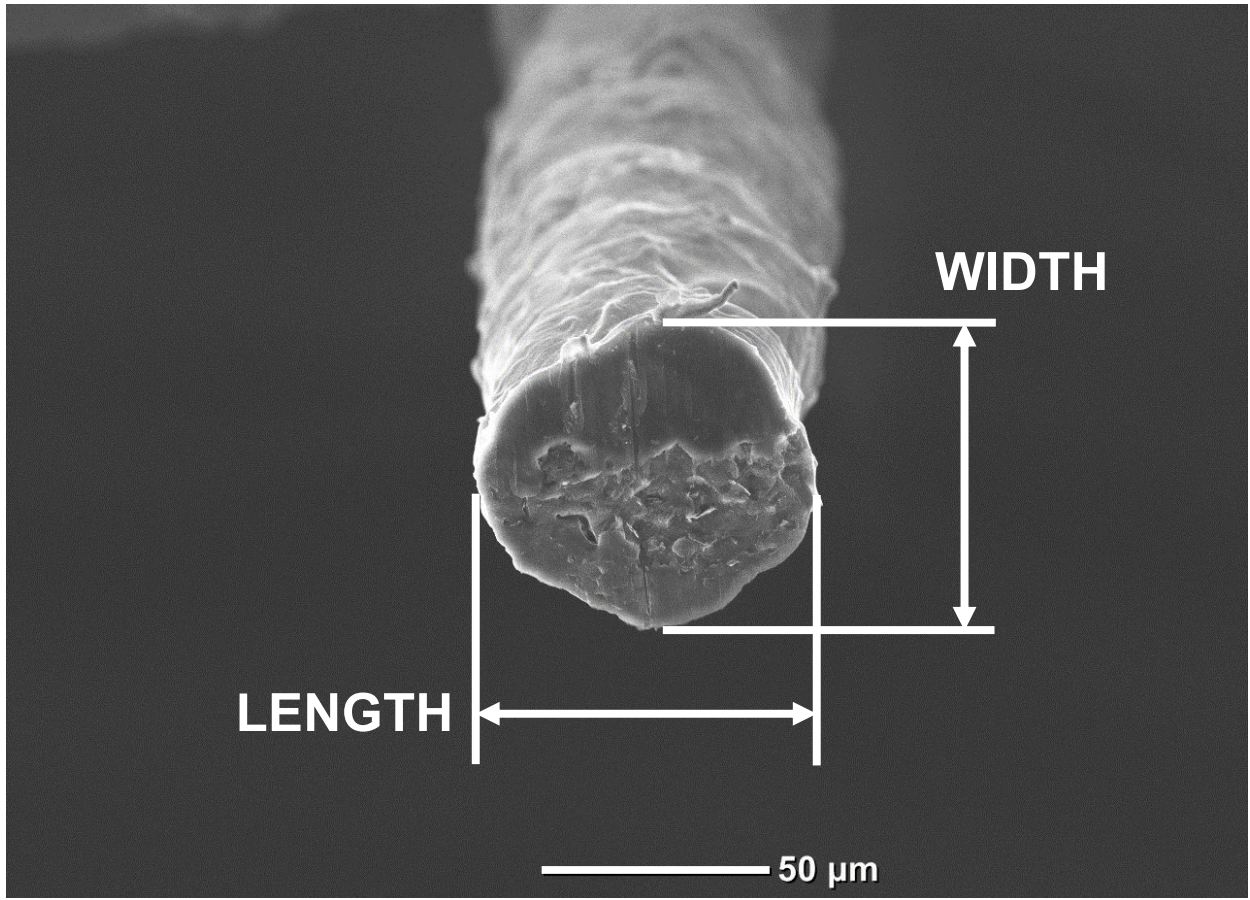

**Figure 6.** Example of fiber cross section defining estimated length and width used as performance parameters for microfiber manufacturing.

**Table 2.** Solid alginate microfiber manufacturing parameters determined experimentally for (a) 0% CaCl$_2$ solution bath and (b) 5% CaCl$_2$ solution bath.

(a)

| Feature | FRR | Mean | Std. Dev. |
|---|---|---|---|
| Porosity (%) | 125:15 | 93.8 | 19.8 |
|  | 100:10 | 22.4 | 2.41 |
|  | 125:10 | 51.6 | 18.3 |
| Fiber Length (µm) | 125:15 | 24.8 | 1.98 |
|  | 100:10 | 16.7 | 3.44 |
|  | 125:10 | 20.0 | 1.36 |
| Fiber Width (µm) | 125:15 | 19.5 | 1.38 |
|  | 100:10 | 14.4 | 1.70 |
|  | 125:10 | 16.9 | 1.27 |
| Young's Modulus (MPa) | 125:15 | 1,750 | 375 |
|  | 100:10 | 402 | 114 |
|  | 125:10 | 1,270 | 303 |

(b)

| Feature | FRR | Mean | Std. Dev. |
|---|---|---|---|
| Porosity (%) | 125:15 | 76.3 | 9.47 |
|  | 100:10 | 12.2 | 2.49 |
|  | 125:10 | 19.0 | 6.40 |
| Fiber Length (µm) | 125:15 | 21.2 | 1.19 |
|  | 100:10 | 7.86 | 1.29 |
|  | 125:10 | 10.3 | 1.86 |
| Fiber Width (µm) | 125:15 | 20.6 | 1.86 |
|  | 100:10 | 6.51 | 0.991 |
|  | 125:10 | 8.24 | 1.34 |
| Young's Modulus (MPa) | 125:15 | 6,010 | 2,300 |
|  | 100:10 | 15,900 | 6,230 |
|  | 125:10 | 8,560 | 1,460 |

Due to sparse availability of raw data, synthetic datasets were generated to improve the training of the DNN. These datasets assume the original measurements follow a Gaussian distribution and therefore carry the mean and standard deviation characteristics of the data reported by [14]. In all, 1,200 datasets covering all combinations of flow rate ratio and bath solution were constructed. Arbitrarily, 479 of these datasets were used for training and testing of the DNN-based model, while 721 were used for independently assessing model accuracy.

### 2.2.3 Model Implementation

The DNN implemented for both predictive and design model generation was constructed to have four dense layers and an output layer. The four dense layers used the rectified linear unit (ReLU) activation function, while the output layer used a linear function. The full set of DNN parameters employed for this study are provided in Table 3.

**Table 3.** Parameters used in implementation of deep neural network for predictive and design model generation.

| Parameter | Value/Type |
|---|---|
| Dense Layers | 4 |
| Learning Rate | 0.001 |
| Epochs | 32 |
| Batch Size | 20 |
| Neurons | 14 |
| Seed Type | Random |
| Activation Function | ReLU |

To observe any improvements in modeling power due to number of batches used, results are presented for batch sizes of 1 through 20. Training and validation loss were monitored over the span of epochs for each consecutive batch in order to observe any overfitting or underfitting of the data during the model generation process.

### 2.2.4 Model Performance Assessment: Error

For the predictive model, accuracy is assessed by comparing the calculated value, $P_P$, of a particular output/performance parameter (cross-sectional length and width, porosity, Young's modulus) with the average value from the test data ($P_T$), holding manufacturing input parameters

(sheath and core flow rate, percentage bath solution) constant. This is expressed as a percentage error:

$$Error = \frac{P_P - P_T}{P_T} \times 100\% \tag{1}$$

For the design model, the same method was used to evaluate sheath and core flow rates for a given desired performance parameter set. Since percentage bath solution for the data was binary (only 0% and 5% available), the results for that manufacturing parameter are simply presented in their basic form.

## 3 Results and Discussion

### 3.1 Predictive Model

Results of predictive model implementation show a large degree of correlation between average values of input parameters and the predicted values based on known inputs. As expected, accuracy of the predictions tends to increase as more batches are included in the resulting model (see Figure 7). However, although the majority of manufacturing input characteristics track closely with empirical values, two outliers are evident from the data. For one, the DNN-based model prediction for a flow rate ratio of 125:10 with a 0% bath solution results in a 6.2% porosity prediction error, which is three times that for the other solutions. Similarly, fiber Young's modulus predictions for a flow rate ratio of 100:10 with a 0% bath solution exceed 30%, which is greater than twice the next highest error.

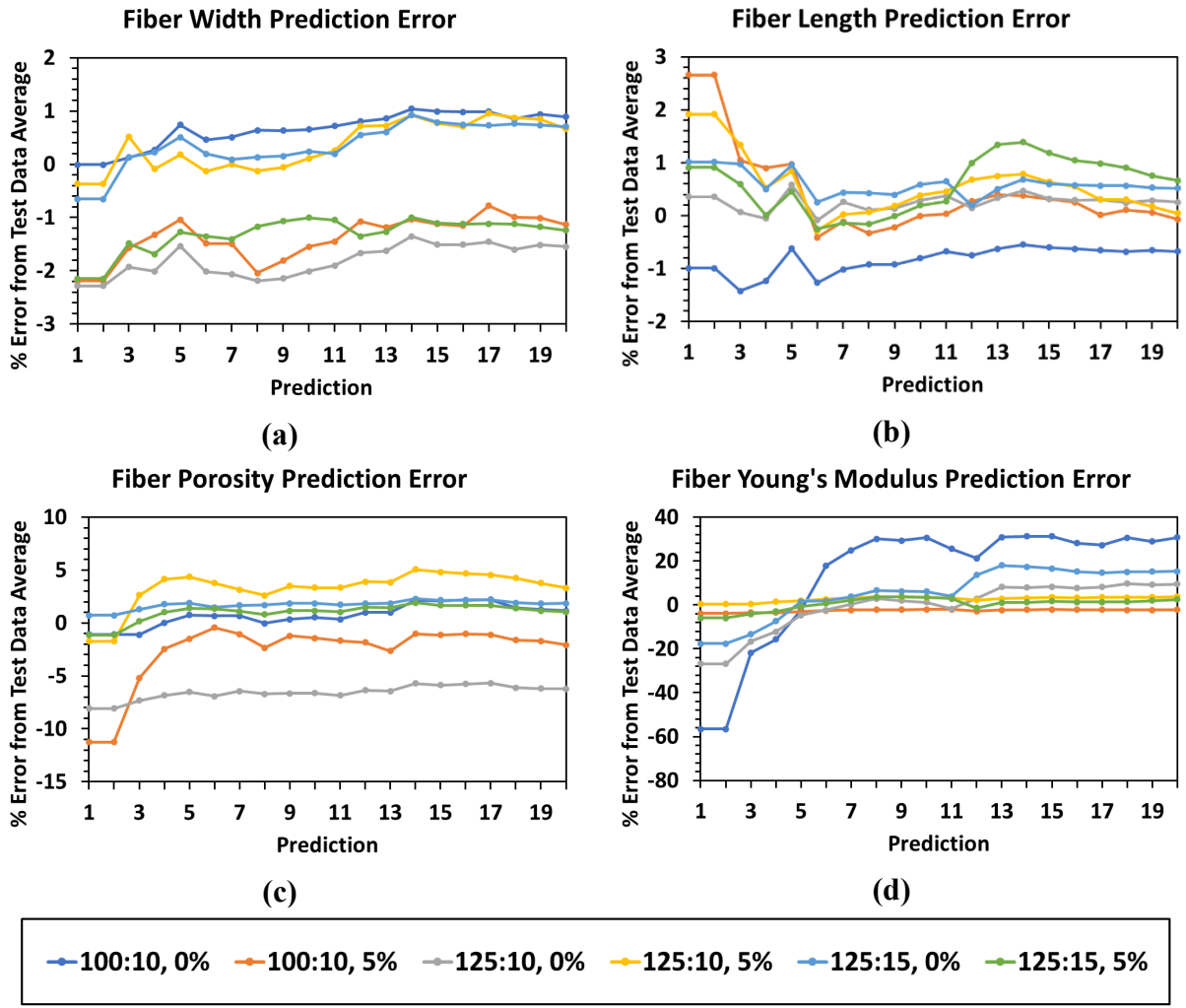

**Figure 7.** Mean prediction error for fiber features calculated using a DNN-based predictive model. (a) cross-sectional width, (b) cross-sectional length, (c) porosity, and (d) Young's modulus. The moniker *prediction* indicates the number of batches combined into the DNN.

One concern with implementation of a DNN using sparse data is overfitting. With the injection of synthetic data generated based on statistic parameters of the original, the hope is that overfitting can be avoided. To assess this, the training and validation loss were evaluated for various batch combinations. As evident from Figure 8, the training and validation loss curves track each other well without eventual uptick of the latter, indicating that overfitting has not, in fact, occurred.

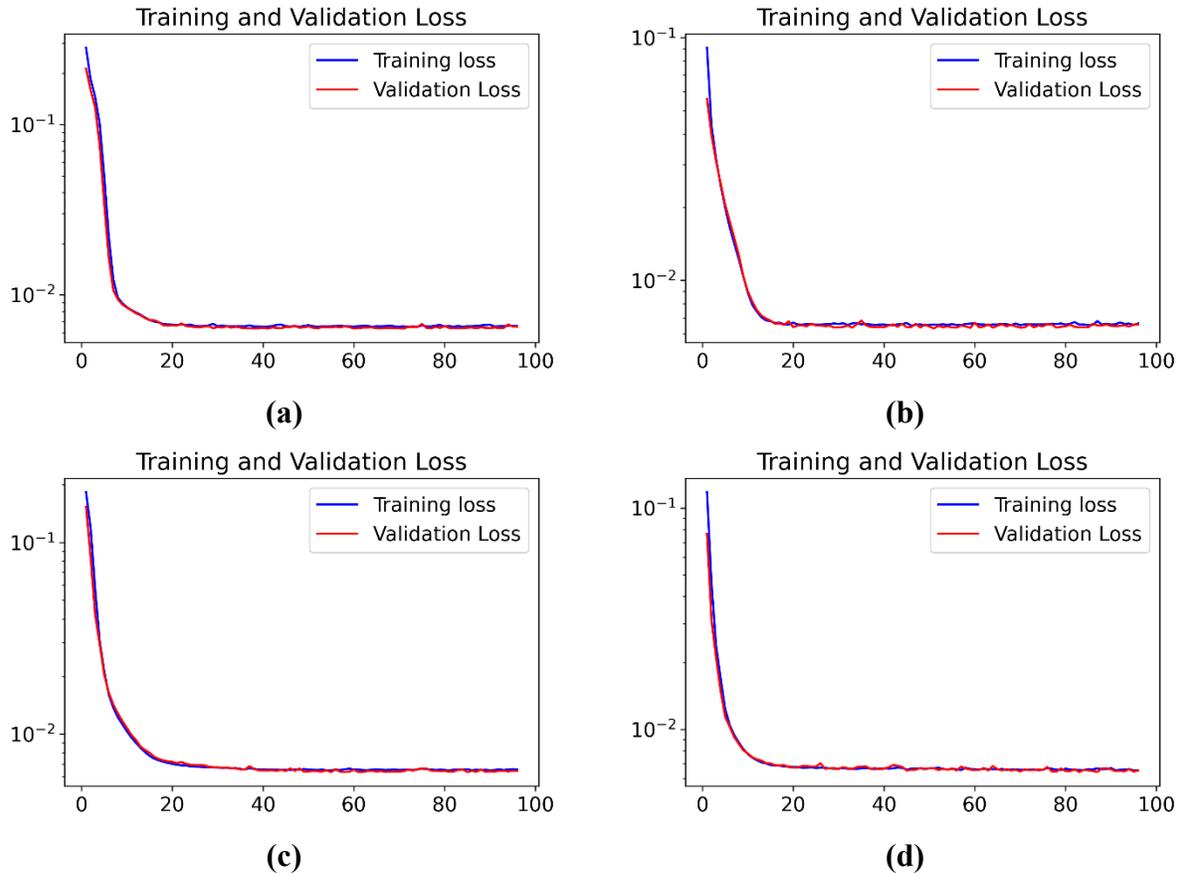

**Figure 8.** Training and validation loss curves for DNN used to generate the predictive microfiber model. Plots represent batch sizes of (a) one, (b) five, (c) ten, and (d) twenty. Note a high correspondence between the two curves all batch sizes, indicating overfitting has not occurred.

## 3.2 Design Model

As with the predictive model, the design model is able to calculate manufacturing parameter values that result in small deviation from the original. Specifically, note the results in Figure 9 show a deviation of less than 3.2% overall for calculated sheath flow rate and 4.5% for core flow rate in comparison to the average of the empirically-based synthesized data. This implies that the DNN-based design model is able to fairly accurately recover average manufacturing parameter values based on desired performance characteristics. This does not, however, signify any level of generalization for the results. Once again, overfitting must be considered as a possible explanation. However, as Figure 10 demonstrates, tight coupling behavior of training and validation loss indicates that overfitting may not be a factor, particularly for models developed using larger batch sizes. Further testing of the data using new datasets not part of (or synthesized from) the original will be useful in further assessing the accuracy and general behavior of the DNN-based model.

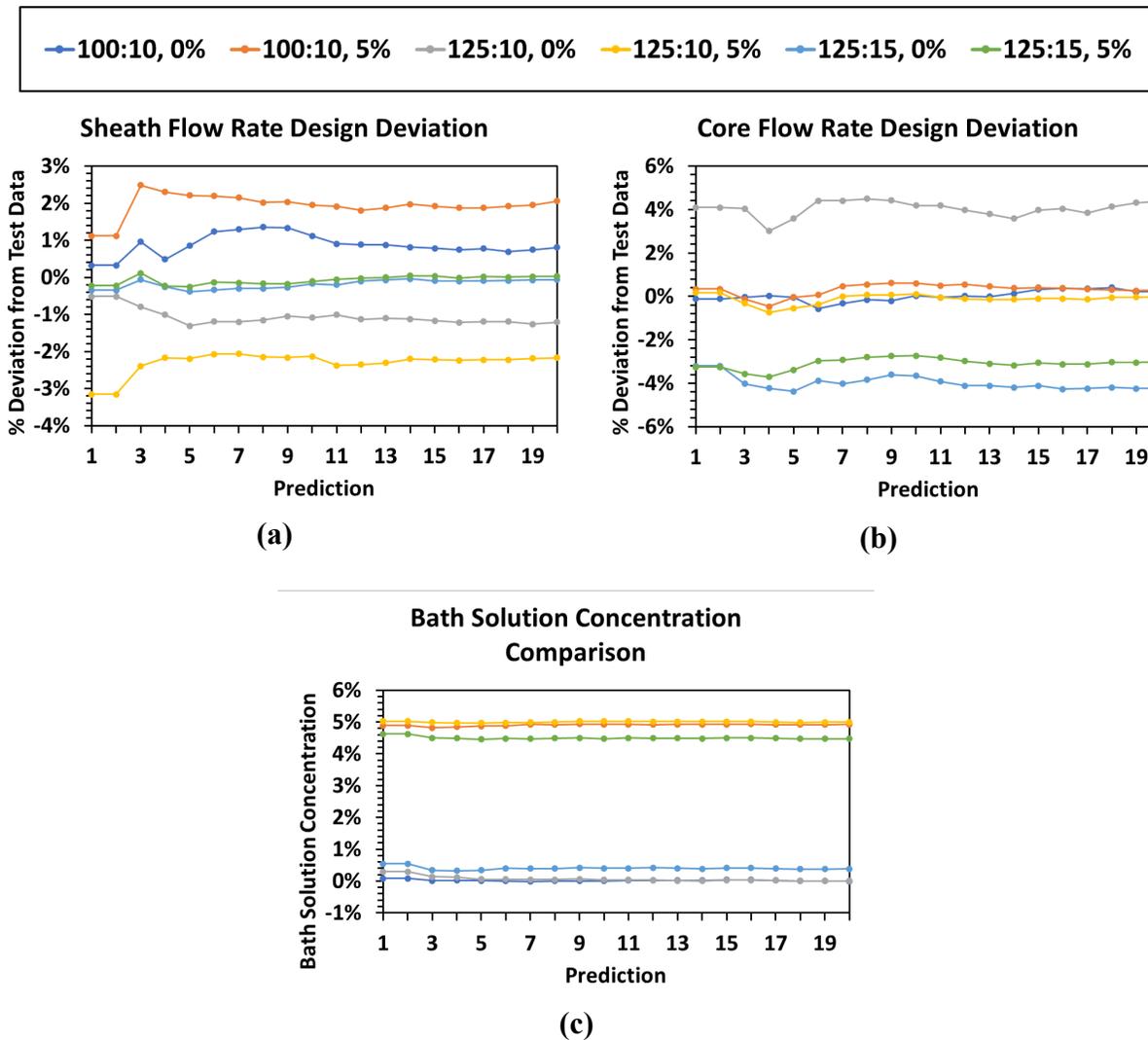

**Figure 9.** Prediction deviation from test data for fiber manufacturing parameters designed using a DNN-based model: (a) sheath flow rate and (b) core flow rate. (c) presents the bath solution selected by the model. Once again, the designation *prediction* indicates the number of batches combined to determine the value of a particular parameter.

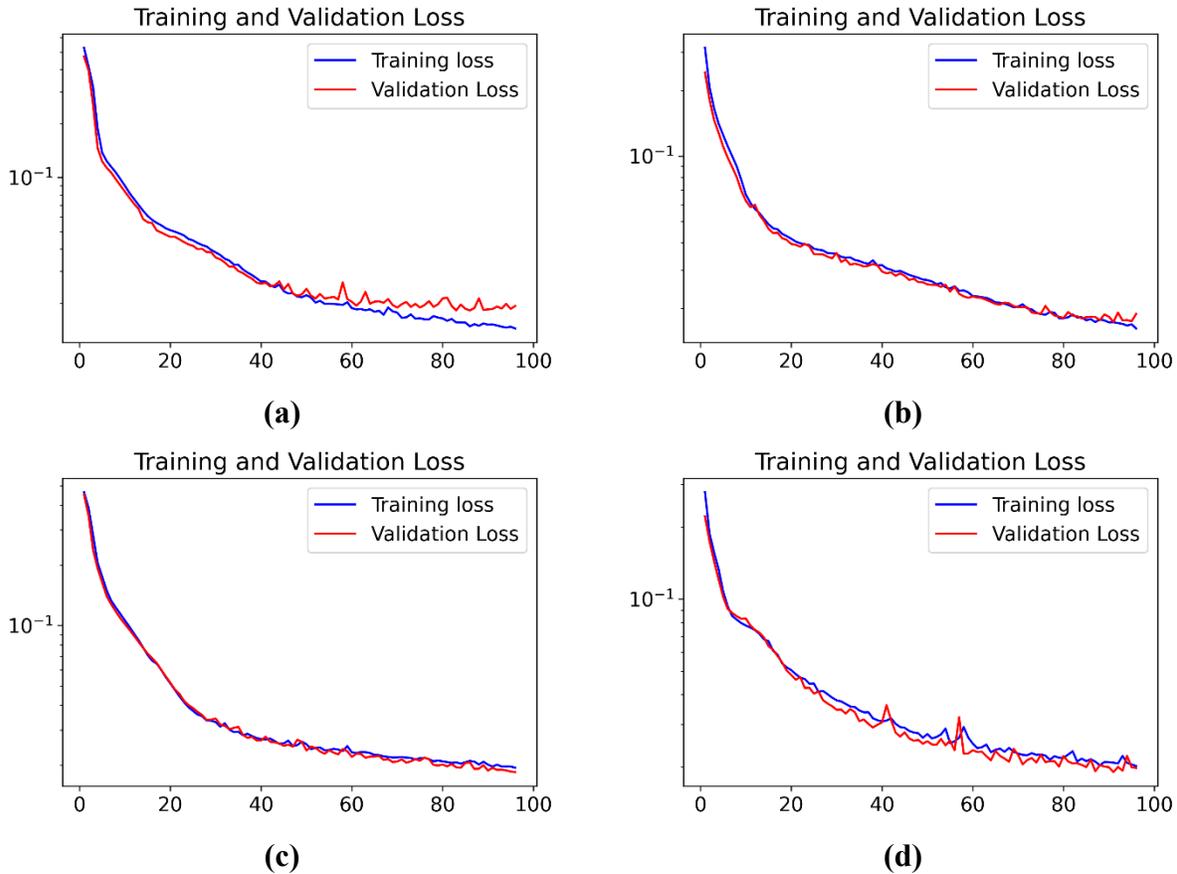

**Figure 10.** Training and validation loss curves for DNN used to generate the microfiber design model. Plots represent batch sizes of (a) one, (b) five, (c) ten, and (d) twenty. Once again, high correspondence between the two curves for larger batch sizes indicates overfitting has not occurred, although for the first two cases a slight uptick in validation loss may be a sign of overfitting. The continual downward trend on (c) and (d) may indicate that full convergence has not quite occurred.

## 4 Conclusions

Results from this study show that DNN-based fiber characteristic prediction and design models have the potential to reduce the amount of trial-and-error currently required with experimental setups. Particularly, design models developed from DNN architectures can potentially speed up the fiber production process by identifying manufacturing parameters given desired performance characteristics. This capability is evidenced by the strong correlation of predicted parameters to the available data.

Where data is sparse or more desired for model robustness, augmentation of training and testing datasets with synthetically-derived values based on statistical distribution of experimental data is a practice that can feasibly increase the viability of DNNs for creating practical predictive and design models to use in microfiber fabrication. Data generated through the method described retained the statistical properties of the original. However, one limitation to the accuracy of results based on synthesized data is worth mentioning. While data was generated using the assumption that they follow a Gaussian distribution, the process described in Section 2.2.2 was conducted independently for each parameter. As such, no correlation among standard deviations for different parameters was made for the synthetic input. As more data is collected and further models refined, this presents an opportunity for model improvement. Additional areas of investigation for enhancement of model predictive and design capability include integration of microfiber scanning electron microscopy (SEM) image processing and integration of physics-based relationships into the DNN code.

**Acknowledgments**

This work was partially supported by the National Science Foundation, Directorate for Technology, Innovation and Partnerships grants 2014346 and 2321975.

**Conflict of Interest**

The authors declare no conflict of interest.

# References

[1] Ouyang L, Armstrong JPK, Salmeron-Sanchez M, Stevens MM. Assembling Living Building Blocks to Engineer Complex Tissues. *Advanced Functional Materials*. 2020;30(26):1909009.

[2] Zaeri A, Zgeib R, Cao K, Zhang F, Chang RC. Numerical analysis on the effects of microfluidic-based bioprinting parameters on the microfiber geometrical outcomes. *Scientific Reports*. 2022;12(1):3364.

[3] Bianco, MJ, Gerstoft, P, Traer, J, Ozanich, E, Roch, MA., Gannot, S, Deledalle, C-A. (2019). Machine learning in acoustics: Theory and applications. *J. Acoust. Soc. Am.* 2019;146(5):3590-3628. doi:10.1121/1.513394

[4] Hashemi, N, Clark, NN. Artificial neural network as a predictive tool for emissions from heavy-duty diesel vehicles in Southern California. *International Journal of Engine Research*. 2007;8(4):321-336. doi:10.1243/14680874jer00807

[5] Frank, M, Drikakis, D., Charissis, V. Machine-learning methods for computational science and engineering. *Computation* 2020;8:15. doi:10.3390/computation8010015

[6] Deelman E, Mandal A, Jiang M, Sakellariou R. The role of machine learning in scientific workflows. *The International Journal of High Performance Computing Applications*. 2019;33(6):1128-1139. doi:10.1177/1094342019852127

[7] Carleo G, Cirac I, Cranmer K, Daudet L, Shuld M, Tishby N, Vogt-Maranto L, Zdeborová L. Machine learning and the physical sciences. *Rev. Mod. Phys*. 2019;91(4): 0450021-31. doi:10.1103/RevModPhys.91.045002

[8] Karpatne A, Khandelwal A, Boriah S, Kumar V. Predictive learning in the presence of heterogeneity and limited training data. *2014 SIAM International Conference on Data Mining*. 2014;1:253-261. doi: 10.1137/1.9781611973440.29

[9] Nguyen LT, Zeng M, Tague P, Zhang J. Recognizing new activities with limited training data. *Proceedings of the 2015 ACM International Symposium on Wearable Computers*. 2015;67-74. doi:10.1145/2802083.2808388

[10] Xi Z, Zhao X. An enhanced copula-based method for battery capacity prognosis considering insufficient training data sets. *2018 Institute of Industrial and Systems Engineers Annual Conference and Expo*. 2018;1306-1311.

[11] Islam J, Zhang Y. GAN-based synthetic brain PET image generation. *Brain Inform*. 2020 Mar 30;7(1):3. doi: 10.1186/s40708-020-00104-2

[12] Tucker A, Wang Z, Rotalinti Y. Generating high-fidelity synthetic patient data for assessing machine learning healthcare software. *npj Digit. Med.* 2020;3(147):1-13. doi:10.1038/s41746-020-00353-9

[13]     Mikołajczyk A, Grochowski M. Data augmentation for improving deep learning in image classification problem. *2018 International Interdisciplinary PhD Workshop (IIPhDW)*. 2018;117-122. doi: 10.1109/IIPHDW.2018.8388338

[14]     McNamara MC, Sharifi F, Okuzono J, Montazami R, Hashemi NN. Microfluidic manufacturing of alginate fibers with encapsulated astrocyte cells. *ACS Appl. Bio Mater*. 2019;2(4):1603-1613. doi: 10.1021/acsabm.9b00022

[15]     Sharifi F, Patel BB, Dzuilko AK, Montazami R, Sakaguchi DS, Hashemi N. Polycaprolactone microfibrous scaffolds to navigate neural stem cells. *Biomacromolecules*. 2016;17(10):3287-3297 doi: 10.1021/acs.biomac.6b01028

[16]     Farrokh S, Bai Z, Montazam R, Hashemi N. Mechanical and physical properties of poly(vinyl alcohol) microfibers fabricated by a microfluidic approach. *RSC Advances*. 2016;6:55343-55353. doi:10.1039/c6ra09519d

[17]     McNamara M, Aykar S, Alimoradi N, Niaraki Asli A, Pemathilaka R, Wrede A, Montazami R, Hashemi N. Behavior of neural cells post manufacturing and after prolonged encapsulation within conductive graphene-laden alginate microfibers. *Advanced Biology.* 2021;5(11). doi:11.1002.adbi.202101026

[18]     Aykar S, Alimoradi N, Taghavimehr M, Montazami R, Hashemi N. Microfluidic Seeding of Cells on the Inner Surface of Alginate Hollow Microfibers. *Advanced Healthcare Materials*. 2022; 11(11). doi.org/10.1002/adhm.202102701